\documentclass[a4paper,11pt]{article}
\usepackage{Layout/pos}
\usepackage{bbold}
\usepackage{amsmath}
\usepackage{booktabs}
\usepackage{lineno}
\usepackage{lipsum}
\usepackage{color}
\usepackage{hyperref}
\usepackage[nameinlink]{cleveref}
\usepackage{graphicx}% Include figure files
\usepackage{dcolumn}% Align table columns on decimal point
\usepackage{slashed}
\usepackage{comment}
\usepackage{wrapfig}
\usepackage{adjustbox}

\usepackage[font=small,labelfont=bf,justification=justified]{caption}
\usepackage[font=small,labelfont=bf, justification=centering]{subcaption}

\newcommand{\clqcd}{\texttt{CL\kern-.25em\textsuperscript{2}QCD}}

\newcommand{\red}[1]{\textcolor{red}{#1}}

\usepackage{placeins}

\title{On the phase structure of massless many-flavour QCD with staggered fermions}
\author*[a]{Jan Philipp Klinger}
\author[a,b]{Reinhold Kaiser}
\author[a,b]{Owe Philipsen}
\author[a]{Jonas Schaible}

\affiliation[a]{Institute for Theoretical Physics - Goethe University, \\
Max-von-Laue-Str. 1, 60438 Frankfurt am Main, Germany}
\affiliation[b]{
    John von Neumann Institute for Computing (NIC) at GSI,
    \newline Planckstr. 1, 64291 Darmstadt, Germany
}

\emailAdd{klinger@itp.uni-frankfurt.de}
\emailAdd{kaiser@itp.uni-frankfurt.de}
\emailAdd{philipsen@itp.uni-frankfurt.de}
\emailAdd{jschaible@itp.uni-frankfurt.de}

\abstract{
When the number of massless fermions exceeds a critical value $N_f^*$, QCD enters the conformal window and becomes chirally symmetric already in the vacuum. Determining $N_f^*$ from lattice simulations is challenging, since calculations are performed at finite lattice spacing, quark mass, and temporal lattice size, where both a thermal transition and an unphysical bulk transition obscure the conformal behaviour.
In this work, we present results on the chiral phase boundaries in the bare lattice parameter space $(N_\tau,\;\beta,\;am,\;N_f)$ of unimproved staggered fermions. Our analysis indicates that the chiral transition in continuum QCD is of second order for all $N_f$ up to the onset of the conformal window. By systematically studying the thermal chiral transition and its interplay with the bulk transition, we obtain a coherent picture of the lattice phase structure and suggest how the onset of the conformal window can be identified from simulations performed away from the chiral and continuum limits.}

\FullConference{The 42nd International Symposium on Lattice Field Theory (LATTICE2025)\\
2-8 November 2025\\
Tata Institute of Fundamental Research, Mumbai, India\\}

\begin{document}
\maketitle

\section{Introduction}
This work builds on previous studies \cite{Cuteri_2021,Klinger:2025mU}, where the chiral phase transition was investigated as a function of the number of degenerate light quark flavours and their mass. Our strategy is to map out the \mbox{(pseudo-)critical} hypersurface of the chiral transition in the extended bare lattice parameter space spanned by the gauge coupling $\beta=6/g^2$, the number of flavours $N_f$, the bare quark mass $am$, and the temporal lattice extent $N_\tau$. It was found that the first-order chiral transitions observed on coarse lattices disappear as the lattice spacing is reduced, indicating that continuum QCD exhibits a second-order chiral transition for $N_f = 2-7$ massless quarks. To complete this study for all numbers of flavours with a chiral phase transition, we explore the phase structure for $N_f = 8$. Interestingly, we observe features that are compatible with the onset of conformal window.
We propose that unraveling the phase structure of the lattice bare parameter space in many-flavour QCD allows for meaningful conclusions about chiral QCD in the continuum and offers a novel perspective on identifying the conformal window through its imprint on the physical phase boundaries. Beyond its intrinsic field-theoretic relevance, the conformal window is of broader interest. Near-conformal gauge theories, characterized by slowly running couplings and potentially large anomalous dimensions, play an important role in composite-Higgs scenarios and other strongly coupled extensions of the Standard Model \cite{beyondSM}. Understanding how conformal dynamics is embedded in the lattice phase structure therefore has implications that extend well beyond QCD itself.

%As we will show, unraveling the phase structure of the lattice bare parameter space in many-flavour QCD allows for meaningful conclusions about chiral QCD in the continuumm, thereby deepening our understanding of strongly interacting matter.
%The conformal window at large $N_f$ and zero density offers a unique opportunity to investigate infrared-conformal dynamics in a controlled, sign-problem-free setting.

\vspace{3mm}
\subsubsection*{Simulation details and analysis method}
Our lattice QCD simulations employ the standard Wilson gauge action and unimproved staggered fermions on lattices of size $N_\tau \times N_\sigma^3$. The lattice gauge coupling $\beta$ controls the lattice spacing~$a$ and tunes the temperature via $T=1/[a(\beta)N_\tau]$. At fixed $T$, the lattice spacing is reduced by increasing $N_\tau$. We used the OpenCL lattice QCD framework \clqcd~\cite{cl2qcd}, with job management and monitoring provided by the bash tool \texttt{BaHaMAS}~\cite{bahamas}.

The chiral transitions are studied by the use of the chiral condensate as a (quasi-)order parameter, $\langle \mathcal{O} \rangle =  \langle \bar{\psi} \psi \rangle$, and its distribution is analysed via its generalised cumulants
\begin{align}
    B_n = \frac{ \left\langle \left( \mathcal{O} - \langle \mathcal{O} \rangle \right)^n \right\rangle\;\;\;\;\;}{\left\langle \left( \mathcal{O} - \langle \mathcal{O} \rangle \right)^2 \right\rangle^{n/2}} \;.
\end{align}
For fixed $N_f$ and $N_\tau$ we locate the transition, i.e. (pseudo-)critical $\beta_{pc}$, at several masses $am$  by scanning in the lattice gauge coupling for vanishing skewness, $B_3(\beta_{pc}, am) = 0$. Its order is then determined by a finite size scaling analysis of the kurtosis $B_4$. We identify the kurtosis $B_4(\beta_{pc}, am, N_\sigma)$ on the transition point for three different aspect ratios $N_\sigma/N_\tau \in \{2,3,4\}$. The associated values of the kurtosis at the transition in the thermodynamic limit ($N_\sigma \rightarrow \infty$) are  $B_4=1$ for a first order transition, $B_4= 1.6044(10)$ for a transition in $Z_2$-universality class and $B_4= 3$ for a crossover. For a more detailed explanation of our procedure of determining the order of the thermal transition see \cite{Cuteri_2021, Klinger:2025mU}.

\newpage
\section{The phase structure of lattice QCD with staggered quarks}
In the following, we present an overview of the lattice phase structure of QCD with unimproved staggered fermions, expressed in terms of the bare lattice parameters $\{N_f,\; am,\; N_\tau,\; \beta\}$. Once the full phase structure is understood, it becomes clear which phases and (pseudo-)critical surfaces connect to the continuum. Of particular interest in the present study is the onset of the conformal window. We propose a picture of how the latter is embedded in the bare lattice phase structure, and how it may be identified by taking appropriate limits in the multidimensional parameter space. We begin by discussing the expected phases and phase boundaries of massless QCD with many flavours in the continuum limit, corresponding to $g =  am = aT = 0$. We then explore how the phase diagram is modified as we move away from the chiral continuum limit towards simulable situations by turning on the relevant extensions: first the gauge coupling $g>0$, then a finite fermion mass $am>0$, and finally a non-zero \mbox{temperature $aT>0$}.
%This step-by-step approach allows us to disentangle the effects of each parameter and to systematically map out the resulting phase boundaries in the full bare parameter space.

\begin{comment}
While our ultimate interest lies in the continuum theory, it is conceptually useful to regard the continuum limit as a particular point --- or more precisely, a critical surface --- within this multidimensional bare parameter space.
\end{comment}

\subsection{Continuum QCD (\boldmath $g=0, \; am=0, \;aT=0$)}

\begin{figure}[htp!]
    \centering
   \includegraphics[scale=0.8]{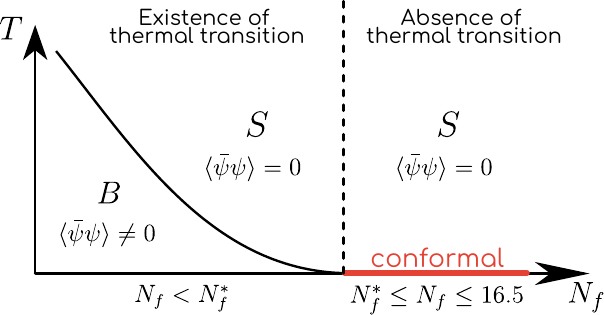}    
    \caption{Chiral phase diagram for massless continuum QCD. Our analysis (see Section \ref{sec: thermal}) suggests a thermal phase transition of second order for all $N_f<N_f^*$ \cite{Cuteri_2021, Klinger:2025mU}.} 
        \label{fig: T_Nf}  
\end{figure}

Massless QCD with many flavours exhibits a finite-temperature phase transition between a chirally broken phase at low temperature and a chirally symmetric phase at high temperature, see Fig.~\ref{fig: T_Nf}. Increasing evidence suggests that this transition is second order for all numbers of flavours~\cite{Fejos_2024, Klinger:2025mU, Cuteri_2021}. For small $N_f$, a nearly linear decrease of the critical temperature with increasing $N_f$ is predicted \cite{Braun:2006jd}. At larger $N_f$, this behaviour transitions into exponential Miransky scaling~\cite{Braun_Fischer, Miransky}, and ultimately the thermal phase transition disappears at the onset of the conformal window. The conformal window is characterized by the cessation of chiral symmetry breaking due to the emergence of a non-trivial infrared (Banks--Zaks) fixed point~\cite{BANKS}. At this fixed point, the running of the coupling stops due to a vanishing beta function, rendering the theory scale invariant. As a result, no thermal phase transition associated with chiral symmetry breaking is present. The perturbative two-loop beta function predicts the appearance of an infrared fixed point for $8.05 \lesssim N_f \leq 16.5$. However, non-perturbative dynamics may significantly modify the lower edge $N_f^*$ of the conformal window, leaving its precise location an open question, see \cite{Kotov:2021hri} for an overview. Recent studies suggest that $N_f=8$ may mark the onset of the conformal window~\cite{Hasenfratz_8, Hasenfratz:2024zqn}. For $N_f > 16.5$, asymptotic freedom is lost as the Gaussian fixed point at $g=0$ ceases to be ultraviolet-attractive. The theory then becomes infrared free, analogous to QED.

%For $N_f > 16.5$, asymptotic freedom is lost, as the one-loop coefficient of the beta function changes sign and the Gaussian fixed point at $g=0$ ceases to be ultraviolet-attractive. The theory then becomes infrared free, analogous to QED.

\begin{comment}
Current studies suggest $
8 \lesssim N_f^* \lesssim 12$ \cite{Hasenfratz_10, Hasenfratz_12, Lombardo_12, Lombardo_8, Kotov:2021hri, Miura:2011mc, Lombardo, Braun:2006jd, Braun:2009ns}, with increasing evidence that $N_f^* \simeq 8$ may mark the sill of the conformal window \cite{Hasenfratz_8, Hasenfratz:2024zqn}.
\end{comment}

\newpage
\FloatBarrier
\subsection{Lattice QCD (\boldmath $g>0, \; am=0, \;aT=0$)}
\begin{figure}[!htp]
   \begin{subfigure}[t]{0.36\textwidth}
    \centering
   \includegraphics[width = \textwidth]{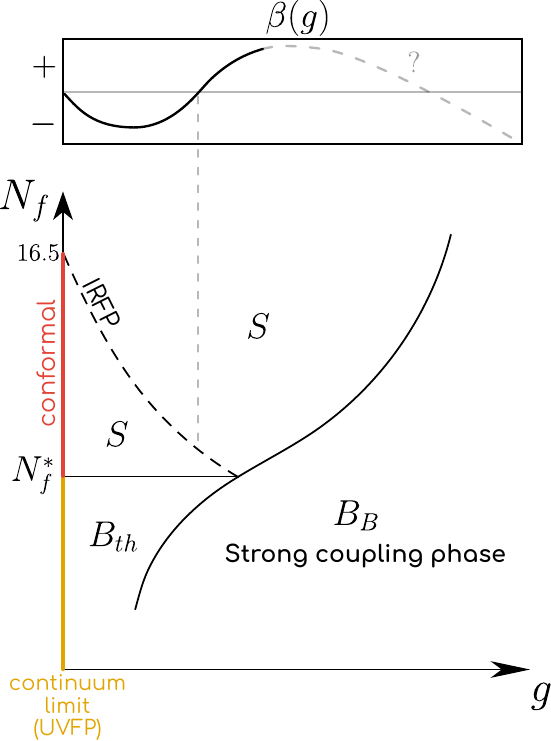}    
    \caption{Projection to $T=0$}
    \label{fig: Mirsanky Nf g}
        \end{subfigure}% 
        \hfill
     \begin{subfigure}[t]{0.55\textwidth}
    \centering
   \raisebox{0.3cm}{ \includegraphics[width = \textwidth]{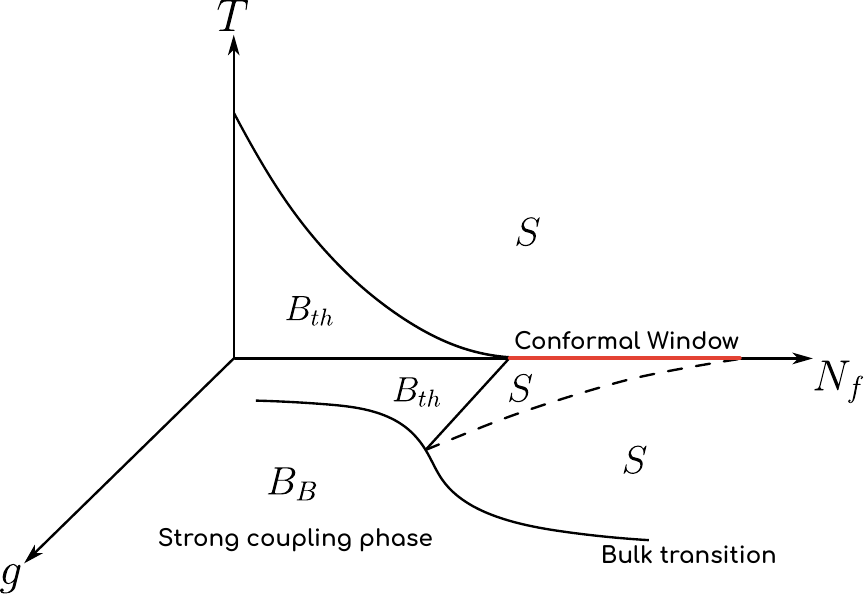}    } 
    \caption{Projection to $aT=0$}
    \label{fig: Miranksy T Nf g}  
    \end{subfigure}
    \caption{Chiral phase diagram for massless QCD at non-zero coupling \cite{Miransky}. A zero-temperature bulk transition separates the $(N_f,g)$-plane into a continuum-connected weak-coupling regime and a strong-coupling regime. The weak-coupling regime divides into a symmetric phase ($S$) and a thermally broken phase ($B_\text{th}$), whereas the bulk regime is chirally broken ($B_B$) by lattice artifacts.} 
    \label{fig: Miransky}
\end{figure}

Turning on a non-zero bare coupling $g>0$ moves the lattice theory away from the continuum limit. Keeping $aT=0$ fixed corresponds to a projection onto the zero-temperature plane. The continuum limit -- and with it the possibility of a genuine thermal phase transition at $aT=0$ -- is recovered only in the limit $g \to 0$. The qualitative structure of the zero-temperature plane at non-zero coupling (Fig.~\ref{fig: Miransky}) was first proposed in \cite{BANKS} and later refined in \cite{Miransky, Appelquist, unger}, see also \cite{Lombardo_12} for an overview. Early mean-field arguments suggested that at sufficiently strong coupling chiral symmetry must be broken for all $N_f$. This implies the existence of a phase boundary separating a physical weak-coupling regime, continuously connected to the continuum, from a strong-coupling regime dominated by lattice artifacts. This zero-temperature bulk transition is driven solely by the gauge coupling and is therefore independent of $N_\tau$. Its shape reflects fermionic colour screening effects: increasing the number of flavours weakens the effective strength of the gauge interaction, and consequently shifts the transition to stronger couplings. Later numerical studies with unimproved staggered fermions showed that chiral symmetry is not broken at strong coupling for all $N_f$. Instead, the line of bulk transitions terminates at $N_f \approx 52$ in the strong coupling limit (not shown) \cite{unger}. Within the weak-coupling region at zero temperature, the theory exhibits a quantum phase transition at a critical flavour number $N_f^*$ (horizontal line in Fig.~\ref{fig: Mirsanky Nf g}). For $N_f < N_f^*$, chiral symmetry is broken at $T=0$ ($B_\text{th}$-phase). For $N_f \geq N_f^*$, chiral symmetry breaking is absent and infrared fixed points (dashed line) emerge making the theory infrared conformal.  The line of fixed points marks the location of zeros of the beta function and separates two chirally symmetric regimes that share the same long-distance behaviour, governed by the infrared fixed points, while differing in their ultraviolet properties.

% On the weak-coupling side, the theory is asymptotically free (quasi-conformal phase) and controlled in the UV by the Gaussian fixed point. On the strong-coupling side, asymptotic freedom is lost and the theory becomes Coulomb-like.

\begin{comment}
Coulomb-like, characterized by a positive beta function and the absence of a Gaussian ultraviolet fixed point. We note that in the quasi-conformal regime, conformal symmetry is not exact at finite scales, since perturbative contributions induce a running coupling. The nature of this running differs on the two sides of the symmetric regimes, reflecting their distinct ultraviolet structures.
\end{comment}

\FloatBarrier

\subsection{Lattice QCD (\boldmath $g>0, \; am>0, \;aT=0$)}
\begin{figure}[!h]
   \begin{subfigure}[t]{0.5\textwidth}
    \centering
\includegraphics[width = \textwidth]{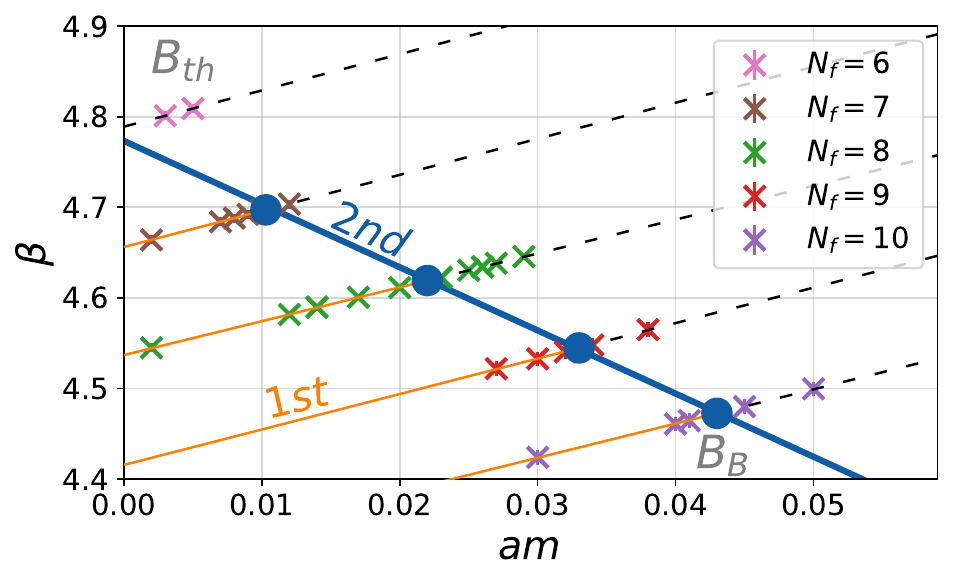}
    \caption{(Pseudo-)critical couplings for $N_\tau \geq 8$}
    \label{fig: Results beta am bulk}
        \end{subfigure}%
     \begin{subfigure}[t]{0.5\textwidth}
    \centering
   \includegraphics[width = \textwidth]{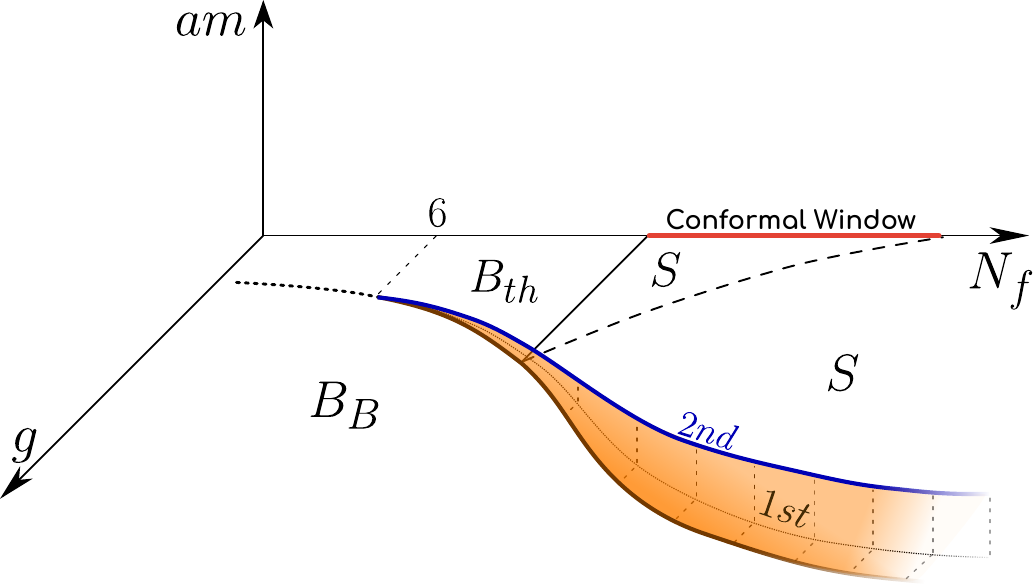}      
    \caption{Projection to $aT=0$}
    \label{fig: illustration bulk}  
    \end{subfigure}
    \caption{Bulk transition between weak- and strong-coupling phases for different $N_f$. The right panel is the extension of Fig.~\ref{fig: Mirsanky Nf g} towards finite bare mass $am$. The bulk transition appears to be non-analytic for $N_f>6$.} 
    \label{fig: results bulk transition total}  
\end{figure}
%1st order transitions are marked in orange, 2nd order in blue and crossover are the dotted lines.

We remain at $aT=0$ and finite coupling $g>0$, thereby continuing to explore the zero-temperature plane while now increasing the bare quark mass $am>0$. In this setup, only a bulk transition can occur. A bulk transition at non-zero bare quark mass was already observed with unimproved staggered fermions in early simulations at $N_f=8$ \cite{christ}. However, to our knowledge, no study has systematically mapped out the bulk transition as a function of both $N_f$ and the bare mass $am$.
Figure~\ref{fig: Results beta am bulk} shows the critical lattice coupling $\beta_c$ and the order of the bulk transition as a function of the bare mass for $6 \leq N_f \leq 10$. The bulk transition separates the weak-coupling regime at large $\beta$ -- which at $aT=0$ and $am>0$ corresponds to the phase $B_{\text{th}}$ for all $N_f$ -- from the bulk regime $B_B$ at small $\beta$. We explicitly verified the bulk nature by comparing results at $N_\tau=8$ and $N_\tau=10$, where the critical couplings $\beta_c(am)$ were found to coincide. For each $N_f$, we observe a linear increase of $\beta_{c}$ on $am$, with a common slope across all flavour numbers.  Increasing $N_f$ shifts the transition to smaller $\beta$, reflecting the enhanced fermionic screening effects discussed in the previous section. Only for $N_f \geq 7$, a first-order bulk transition is found at small masses, which eventually terminates at a second-order critical point and turns into a crossover at larger masses. Further, it was found that the second-order points for different $N_f$ align along a straight line in the $(\beta,am)$ projection. When extrapolated to $am=0$, this line intersects the mapped-out crossover line for $N_f=6$ approximately in the chiral limit. This observation suggests that a non-analytic bulk transition emerges only for $N_f>6$, while for $N_f\lesssim6$ the bulk transition is analytic and manifests itself merely as a crossover. A similar behaviour has been reported for Wilson fermions \cite{wilson_bulk}.

The resulting qualitative picture is summarised in Fig.~\ref{fig: illustration bulk}, which displays the phase boundaries in the zero-temperature plane. The first-order bulk transition, present in the chiral limit for $N_f>6$, extends to non-zero masses and is bounded by a second-order line. As the fermion mass increases, the (pseudo-)critical coupling shifts toward smaller bare coupling $g$. As before, this behaviour can be understood in terms of the effective strength of the gauge interaction. Increasing the fermion mass suppresses dynamical fermion fluctuations and thereby reduces screening. %i.e., increasing $am$ induces the same effect as a reduction of $N_f$,
%namely stronger effective interactions.
%fewer active fermionic degrees of freedom imply weaker screening and thus 
After all, varying any of $g$, $N_f$ or $am$  effectively tunes the interaction strength, which determines the phase of the theory at $aT=0$.

%This enhances the effective gauge interaction and drives the system toward the strongly coupled regime. 

\begin{comment}
As before, this behaviour can be summarised in terms of the effective strength of the gauge interaction. Increasing the number of flavours enhances fermionic screening, which weakens the effective gauge interaction and favors the weakly coupled phase. Conversely, increasing the bare coupling strengthens the interaction and drives the system toward the strongly coupled phase. Similarly, increasing the fermion mass suppresses dynamical fermion fluctuations and reduces screening. This effectively enhances the gauge interaction, pushing the system toward the strongly coupled regime. In this sense, increasing $am$ qualitatively induces the same effect like reduction of $N_f$: fewer active fermionic degrees of freedom lead to weaker screening and therefore stronger effective interactions.
\end{comment}

\FloatBarrier
\begin{comment}
When a quark mass is heavy in the sense that it is much larger than the inverse of the lattice spacing, the effect of quark loops should be negligible and therefore the system should belong to the same universal class as the quenched QCD. Thus the quark should be confined. When the quark mass becomes smaller, the effect of quark loops becomes crucial and there is a possibility that the system enters into a new phase because of this effect. When Nf is as large as 18, this effect indeed triggers the transition observed above.
\end{comment}

\newpage

\subsection{Lattice QCD (\boldmath $g>0, \; am>0, \;aT>0$)}
\label{sec: thermal}
Finally, we turn to lattice QCD at finite temperature, $aT>0$, thereby leaving the zero-temperature plane and allowing for the possibility of a genuine thermal phase transition. The introduction of temperature adds an additional phase boundary: the weak-coupling regime, which at $aT=0$ was chirally broken ($B_\text{th}$), can now undergo thermal restoration of chiral symmetry to a symmetric phase $S$. 
The chiral phase boundary associated with the thermal transition was already investigated in previous works \cite{Cuteri_2021, Klinger:2025mU} for $2 \leq N_f \leq 8$. There, the primary focus was on determining the order of the thermal transition in the chiral limit. In agreement with earlier studies on coarse lattices, a first-order transition was observed at small quark masses. However, it was also demonstrated that the size of the first-order region decreases as $N_\tau$ increases
and that, in the lattice chiral limit $am=0$, a first-order transition persists only up to a tricritical lattice spacing, viz.~temporal extent $N_\tau^{\mathrm{tric}}$. For finer lattices, $N_\tau \geq N_\tau^{\mathrm{tric}}$, the chiral transition becomes second order. 
Such a tricritical point was found for all $N_f \leq 7$. For $N_f \geq 8$ the presence of the bulk transition interferes with the thermal critical structure and prevents the realisation of a tricritical point. This behaviour can be linked to the possibility that $N_f=8$ marks the onset of the conformal window. In the following, we first discuss the thermal transition for $N_f \leq 7$ and then turn to $N_f = 8$, where the interplay between thermal and bulk transitions becomes relevant.

\begin{comment}
shorter alternative:
In previous studies of our group it was shown that the regions of first-order chiral transitions,
observed explicitly for Nf = 2 - 7 degenerate light quarks on coarse lattices, cf. Fig. 1 (right),
shrink and disappear each in a tricritical point at vanishing bare quark mass for some finite lattice
spacing in units of temperature, aT 6= 0
\end{comment}

\FloatBarrier
\paragraph{Thermal transition for \boldmath  $N_f \leq 7$:}
\begin{figure}[!b]
\captionsetup[subfigure]{skip=2pt}
\captionsetup{skip=6pt}
\vspace{-1mm}
        \begin{subfigure}[t]{0.6\textwidth}
    \centering
\includegraphics[width = \textwidth]{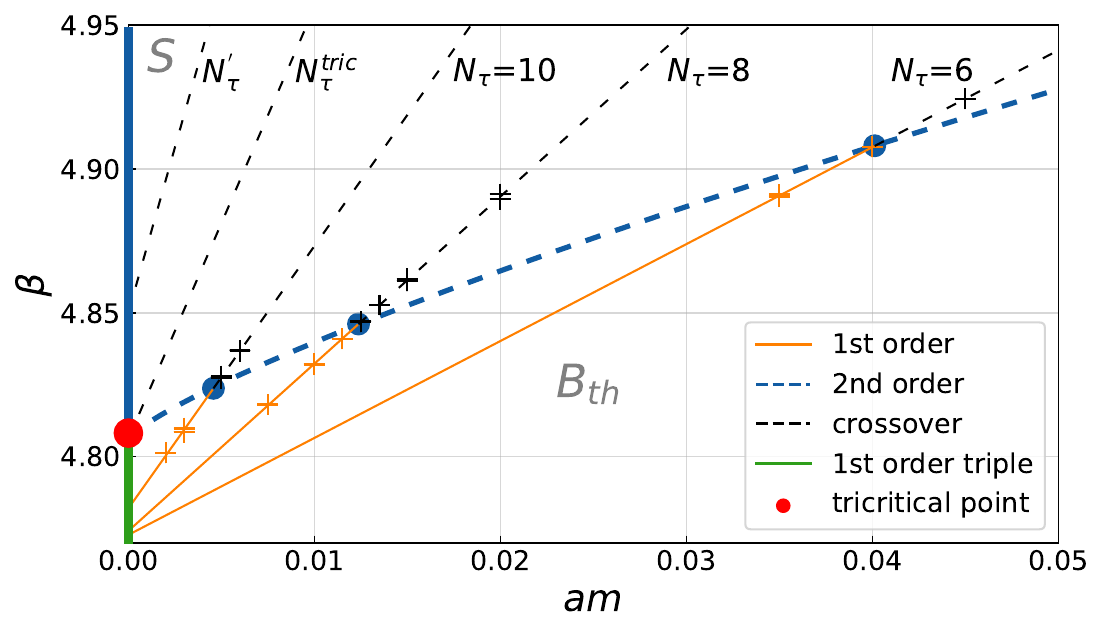}
    \caption{(Pseudo-)critical couplings for $N_f=6$}
    \label{fig: Results Nf6 beta am}  
    \end{subfigure}%
 \begin{subfigure}[t]{0.4\textwidth}
    \centering
\raisebox{5mm}{\includegraphics[width = 0.87\textwidth]{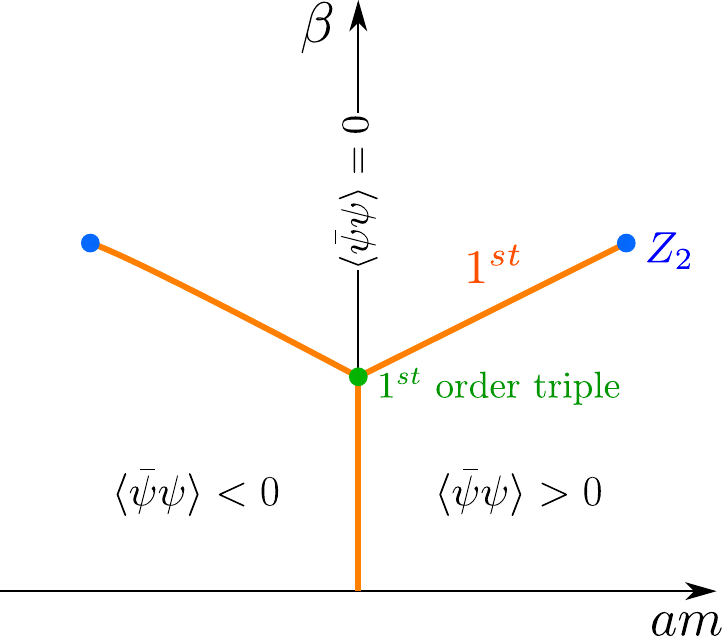}}
    \caption{ 3-state coexistence at fixed $N_\tau$}
    \label{fig: illustration beta am}  
        \end{subfigure}
\caption{ Thermal phase transition for different temporal extent $N_\tau$. The 1st order region is bounded by a 2nd order wing line, which terminates at a tricritical point $(\beta^{\mathrm{tric}}, N_\tau^{\mathrm{tric}})$ as $N_\tau$ increases. For $N_\tau \geq N_\tau^{\mathrm{tric}}$, the thermal transition in the chiral limit is of 2nd order, while for $am>0$ it turns into a crossover. As a consequence, the chiral continuum limit ($am=0$, $\beta \to \infty$, $N_\tau \to \infty$) is governed by a 2nd order transition.} 
    \label{fig: thermal transition}
\end{figure}

In the following we pick $N_f = 6$ as a representative example for all $N_f\leq 7$, which share the same qualitative behaviour. Figure \ref{fig: Results Nf6 beta am} shows the (pseudo-)critical coupling as a function of the bare mass for $N_\tau = 6, 8, 10$. For each $N_\tau$ the weak-coupling regime separates into a symmetric phase $S$ (above the transition line) and broken phase $B_\text{th}$ (below transition line). A first-order thermal transition (orange) was found at small masses for each simulated $N_\tau$. With increasing mass, the first-order line terminates at a second-order critical point (blue) and turns into a crossover (black). Every first-order line originates from a triple point (green) on the y-axis, characterized by the coexistence of three distinct states corresponding to vanishing, positive, and negative values of the chiral condensate (see Fig.~\ref{fig: illustration beta am}). For the first-order transition to be physical, the corresponding triple line (in Fig.~\ref{fig: Results Nf6 beta am}) would have to extend to $\beta \to \infty$ as $N_\tau \to \infty$, in order to connect to the continuum. However, the analyses in \cite{Cuteri_2021, Klinger:2025mU} provide evidence for a tricritical point at $am=0$ for all $N_f \leq 7$. This point $(\beta^{\mathrm{tric}}, N_\tau^{\mathrm{tric}})$ marks the end of the first-order triple line and the onset of a second-order line. From it, a second-order $Z_2$ wing line emerges into the $(\beta,am)$-plane, separating the first-order region (below) from the crossover region (above):
\vspace{-0.5em}
\begin{align}
\label{eq: tric scaling}
    \beta_c(am) = \beta^{\mathrm{tric}} + A (am)^{2/5} + B (am)^{4/5} + \mathcal{O}(am^{6/5}) \, .
\end{align}
\vspace{-2em}\par
\noindent
The dashed blue line in Fig.~\ref{fig: Results Nf6 beta am} is obtained by fitting the 2nd order points for different $N_\tau$ to Eq.~\ref{eq: tric scaling}.
\noindent
In conclusion, the order of the chiral transition of QCD with massless quarks is of 1st order for coarse lattices $N_\tau<N_\tau^{tric}$ and of 2nd order for lattices with $N_\tau\geq N_\tau^{tric}$. The first-order region is merely a cutoff effect, and the transition in the continuum chiral limit is second order for all \(N_f \leq 7\).

\begin{comment}
\red{Even though the first order decreses $am_Z2->0$, it potentially good be that the first order is connected to the continuum and Nt,beta -> infty. For a fixed Nf a tricritical point is not necessary to exist. However in Ref 2021 and Proceeding it was shown that for each Nf a tricritical point exists at $N_tric$,  }
\end{comment}

\FloatBarrier
\paragraph{Thermal transition for \boldmath $N_f=8$:}

\begin{figure}[!b]
    \centering
   \includegraphics[width = 0.6\textwidth]{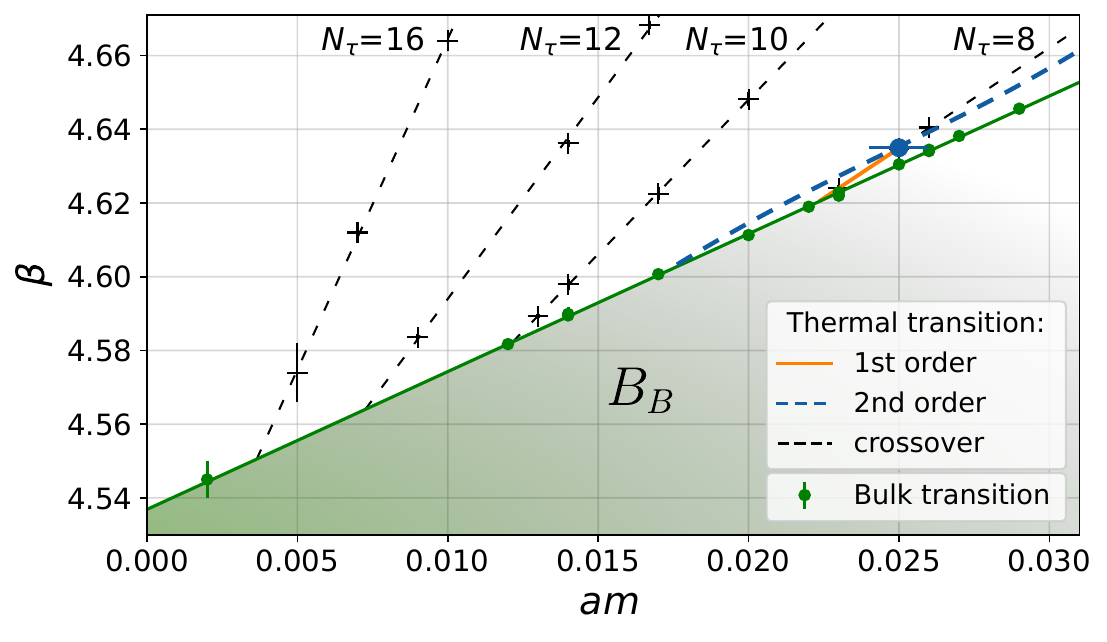}    
    \caption{Thermal and bulk transition for $N_f=8$. Within the weak-coupling regime, the thermal transition depends on $N_\tau$ and separates the symmetric phase $S$ from the thermally broken phase $B_\text{th}$. As $N_\tau$ increases from $N_\tau=8$ to $N_\tau=10$, the first-order thermal transition (orange) vanishes into the bulk regime (green).}
    \label{fig: illustration Nf8 small Nt}  
\end{figure}

At $N_f=8$, the bulk transition starts to interfere with the line of second order transitions (Eq.~\ref{eq: tric scaling}). Figure~\ref{fig: illustration Nf8 small Nt} summarises our results for temporal extents $N_\tau \geq 8$. The bulk transition separates the bulk regime $B_B$ from the physical weak-coupling regime. The latter is again subdivided by the thermal transition into a symmetric phase $S$ and a broken phase $B_{\text{th}}$. While the bulk transition at fixed mass is independent of $N_\tau$ (for $N_\tau \geq 8$), the (pseudo-)critical coupling of the thermal transition shifts with $N_\tau$ analogous to Fig.~\ref{fig: Results Nf6 beta am}. We find that the thermal transition becomes masked by the bulk phase at small $\beta$ for each simulated $N_\tau$. Consequently, the thermal transition for fixed $N_\tau$ persists only at sufficiently large masses. Moreover, only for $N_\tau=8$ do we observe a first-order thermal transition, which then disappears into the bulk transition for decreasing mass. For $N_\tau \geq 10$, neither a first-order transition nor a second-order critical point is found; only a crossover remains. We therefore conclude that for $N_f=8$ the first-order thermal transition is not connected to the chiral limit but instead ends on the bulk transition at finite mass. This contrasts with the case $N_f \leq 7$, where the line of second-order transitions separating the first-order region from the crossover terminates in a tricritical point at $am=0$. For $N_f=8$, by contrast, the $Z_2$ line ends on the bulk transition between $N_\tau=8$ and $N_\tau = 10$. Note that in Fig.~\ref{fig: illustration Nf8 small Nt} the $Z_2$ line (blue) is not based on a fit to Eq.~\ref{eq: tric scaling}, but is shown schematically to illustrate its disappearance into the bulk regime.

\newpage
\paragraph{Conjectured thermal transition for \boldmath $N_f \geq N_f^*$:}
\begin{figure}[htp!]
   \begin{subfigure}[t]{0.5\textwidth}
    \centering
   \includegraphics[width =0.95\textwidth]{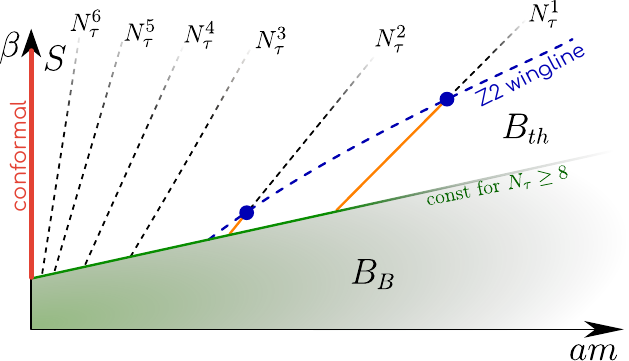}  
    \caption{Scenario if $N_f=8$ is conformal}
    \label{fig: illustration Nf8 conformal}  
        \end{subfigure}%
     \begin{subfigure}[t]{0.5\textwidth}
    \centering
\includegraphics[width =0.95 \textwidth]{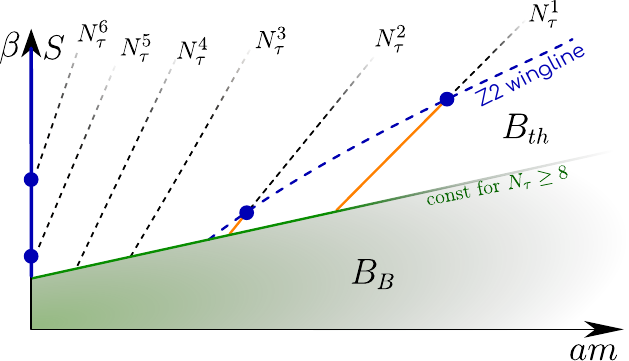}
    \caption{Scenario if $N_f=8$ is non-conformal}
    \label{fig: illustration Nf8 non-conformal}  
    \end{subfigure}
    \caption{Conjecture for the phase structure of $N_f=8$ for large $N_\tau$. If $N_f=8$ is conformal (left), the thermal transition is absent in the chiral limit. In this case, the transition lines at fixed $N_\tau$ intersect the bulk transition at finite mass $am>0$. In contrast, if $N_f=8$ is not conformal (right), a second-order transition in the chiral limit would emerge and connect to the continuum ($\beta \to \infty$ as $N_\tau \to \infty$), similar to Fig.~\ref{fig: Results Nf6 beta am}.}
        \label{fig: illustration Nf8 large Nt}  
\end{figure}

The conformal window is characterised by the absence of a thermal phase transition in the chiral limit and therefore implies chiral symmetry at all temperatures, including $T=0$. This means that in Fig.~\ref{fig: illustration Nf8 small Nt} the lattice chiral limit, $am=0$, must be in the symmetric phase for all $N_\tau$
until $\beta$ is small enough to transition into the bulk phase.
Consequently, the thermal transition lines for increasing $N_\tau$ must intersect the bulk transition at finite mass $am>0$ with increasing slopes, such that no thermal transition can reach the lattice chiral limit.
Figure~\ref{fig: illustration Nf8 conformal} illustrates this conjectured behaviour of the thermal transition at large $N_\tau$. This picture 
recovers that of Fig.~\ref{fig: Mirsanky Nf g} in the limit $am=aT=0$, with only a bulk transition between phases $S$ and $B_\text{B}$ for all $N_\tau$.  The thermally broken phase $B_{\text{th}}$ is then disconnected from the chiral limit in the continuum ($am=0$, $\beta \to \infty$, $N_\tau \to \infty$), as it must be, and the conformal window emerges at $am=0$ as $N_\tau \to \infty$. By contrast, when 
$N_f<N_f^*$ (see Fig.~\ref{fig: illustration Nf8 non-conformal}) the crossover lines for large $N_\tau$ terminate in second order points in the lattice chiral limit, which must be continuously connected to the physical second-order transition in the continuum limit.

Comparing Fig.~\ref{fig: illustration Nf8 conformal} with Fig.~\ref{fig: illustration Nf8 non-conformal}, we propose that the steepening of the thermal crossover lines with increasing $N_\tau$, together with their increasingly dense intersections with the bulk transition, may provide a testable signature to distinguish $N_f > N_f^*$ from $N_f < N_f^*$ in simulations at finite mass and temporal lattice extent. So far, our results are not sufficient to determine whether $N_f=8$ lies within the conformal window. Simulations up to $N_\tau \leq 16$ have not revealed a thermal transition in the chiral limit that is connected to the continuum -- a marked difference compared to all $N_f \leq 7$. In order to fully exclude the scenario in Fig.~\ref{fig: illustration Nf8 non-conformal}, additional simulations at larger $N_\tau$ are required.

\begin{comment}
In the bare lattice phase diagram conformal symmetry is easily lost, since even for $N_f$ inside the conformal window a thermal phase transition must appear once $am>0$. However, the conformal window needs to arise at a vanishing mass for $N_f\geq N_f*$ in the zero-temerpature limit, i.e. $aT \to 0$. In terms of Fig.~\ref{fig: illustration Nf8 small Nt}, this implies that at $am=0$ the chirally symmetric phase of the weak-coupling regime must persist in the limit $N_\tau \to \infty$. The only way this can occur is if all thermal transition lines at fixed $N_\tau$ intersect the bulk transition at finite mass, $am>0$, before reaching the chiral limit. Figure~\ref{fig: illustration Nf8 conformal} shows our conjectured behaviour of the thermal transition lines at large $N_\tau$, under the assumption that $N_f=8$ lies inside (or at the onset of) the conformal window. In order to recover the Miransky picture at $am=aT=0$ limit, the phase diagram at $am=0$ must contain only the bulk transition between $S$ and $B_S$. This requires that the thermal transition is absent in the
\end{comment}

\FloatBarrier
\section{Conclusion}
\begin{figure}[htp!]
   \begin{subfigure}[t]{0.5\textwidth}
    \centering
\includegraphics[width=\textwidth]{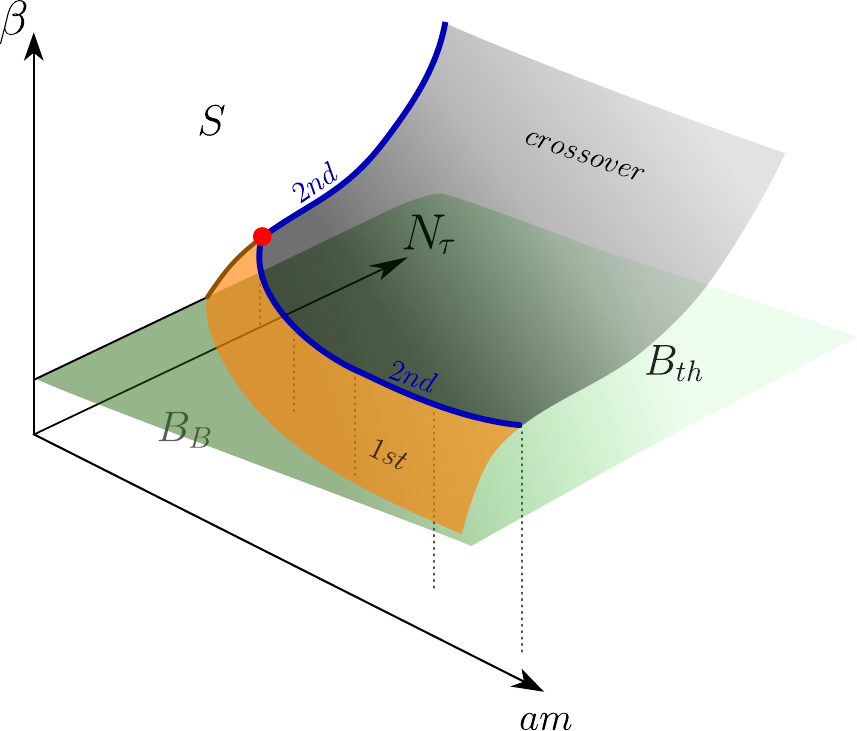}
    \caption{Our results for $N_f \leq 7$}
    \label{fig: 3D beta Nt am nonconformal}  
        \end{subfigure}%
     \begin{subfigure}[t]{0.5\textwidth}
    \centering
\includegraphics[width=\textwidth]{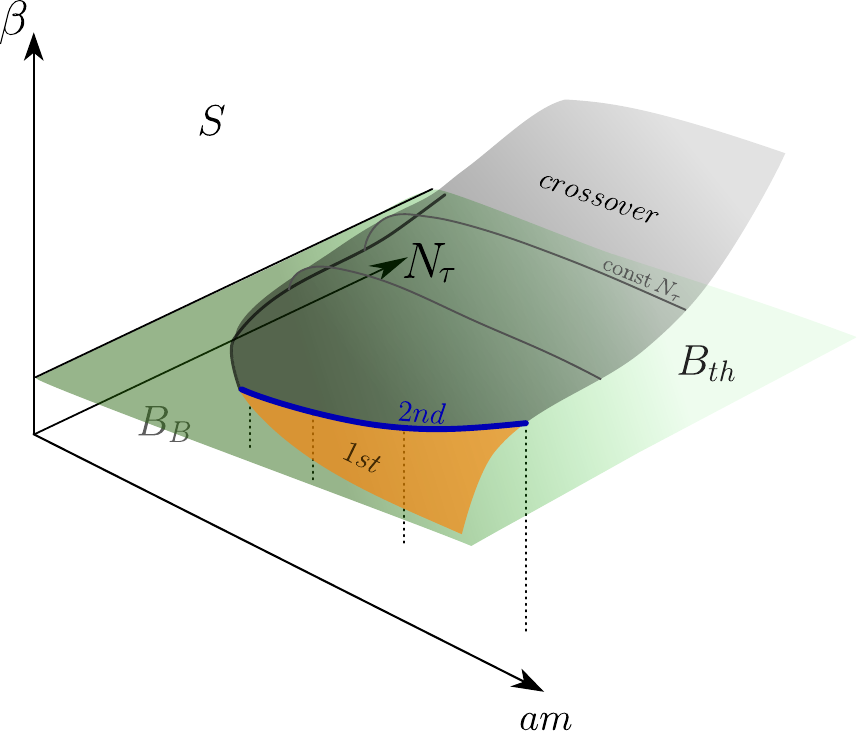}  
    \caption{Conjecture for conformal $N_f$ }
    \label{fig: 3D beta Nt am conformal}  
    \end{subfigure}
    \caption{Full phase structure of the bare parameter space $(N_\tau,\beta,am)$ of lattice QCD for $N_\tau \geq 8$. The left panel illustrates our results for $N_f \leq 7$, where a tricritical point (red) is found. This point marks the change from a 1st order to a 2nd order transition in the chiral limit as $N_\tau$ increases. The right panel shows our conjectured phase structure for conformal $N_f$. In this scenario, the absence of a thermal transition in the massless plane requires the thermal surface to intersect the bulk transition at finite mass $am>0$ for all $N_\tau$.}
        \label{fig: 3D beta Nt am}  
\end{figure}

We summarise our results by presenting the phase structure of lattice QCD with staggered quarks in Fig.~\ref{fig: 3D beta Nt am}, which shows the chiral phase boundaries in terms of $N_\tau$, $\beta$, and $am$. Figure~\ref{fig: 3D beta Nt am nonconformal} corresponds to the case of non-conformal flavour numbers, while Fig.~\ref{fig: 3D beta Nt am conformal} illustrates the scenario for $N_f$ inside the conformal window. Regardless of the flavour number, the phase diagram exhibits a zero-temperature bulk transition separating the physical weak-coupling regime from a bulk regime dominated by lattice artifacts ($B_B$). Its critical coupling is independent of $N_\tau$, increases linearly with~$am$, and is shown as the flat green surface. For $N_f>6$, the bulk transition becomes non-analytic at small masses; for simplicity, its order is not indicated in the figure. Within the physical regime, an additional phase boundary arises from the thermal transition, separating the chirally symmetric phase $S$ from the thermally broken phase $B_{\text{th}}$. On coarse lattices (small $N_\tau$), a first-order thermal transition (orange) is observed for all $N_f$. This first-order region is bounded by a second-order line that terminates at $am=0$ in a tricritical point $(\beta^{\mathrm{tric}}, N_\tau^{\mathrm{tric}})$ for $N_f \leq 7$. Consequently, the chiral transition is first order for $N_\tau < N_\tau^{\mathrm{tric}}$ and second order for $N_\tau \geq N_\tau^{\mathrm{tric}}$.
%With the improved understanding developed since our analysis in \cite{Cuteri_2021}, we estimate the tricritical temporal extent to be around $N_\tau^{\mathrm{tric}} \approx 22$ for all flaovurs.
For $N_f \geq 8$, the presence of the bulk transition interferes with the thermal critical structure and prevents the realisation of a tricritical point. The first-order region does not terminate in the chiral limit, but merges into the bulk regime already at finite mass. 
Unless hitherto unseen first-order transitions show up for larger~$N_\tau$, this
implies second-order phase transitions  also for $N_f>7$ up to the conformal window.

For $N_f \geq N_f^*$, the thermal transition must be absent in the massless plane, see Fig.~\ref{fig: 3D beta Nt am conformal}. In this case, only the bulk transition between the phases $B_B$ and $S$ remains at $am=0$ for all~$N_\tau$, implying that the thermally broken phase $B_{\text{th}}$ is not connected to the chiral continuum limit \mbox{($am = 0,\;\beta \to \infty$,\; $N_\tau \to \infty$)}. 
For this scenario to hold, the thermal and bulk transition surfaces must intersect at finite mass for all $N_\tau$. This observation suggests a strategy to distinguish conformal from non-conformal flavour numbers: for non-conformal $N_f$, the intersection line reaches the chiral limit at finite $N_\tau$, whereas for conformal $N_f$ this would occur only in the limit $N_\tau \to \infty$. We find
$N_f=8$ to be a candidate for this behaviour, but confirming this scenario requires simulations at still larger $N_\tau$.

\begin{comment}
\red{Further, we do not claim to have directly studied the physics around the IRFP itself. The latter type of study is notoriously difficult, while the strategy presented here aims at probing the emergence of conformality in an indirect way.}

\red{In the bare lattice phase diagram, conformal symmetry is easily lost, since even for $N_f$ inside the conformal window a thermal transition appears once $am>0$. }
\end{comment}

%Emphasize more that we use staggered fermions and that whole picture might change ?

\newpage
\renewcommand{\acknowledgments}{%
  \section*{Acknowledgments}
  \vspace{-1em}
}

\acknowledgments{ \noindent The work is supported by the Deutsche Forschungsgemeinschaft (DFG) through the CRC-TR 211 – project number 315477589 – TRR 211. We acknowledge computing resources provided by the VIRGO cluster at GSI Darmstadt and the Goethe-HLR cluster at the Center for Scientific Computing Frankfurt, as well as the use of the unpublished software MC Analysis by J. Schaible and R. Kaiser.}

\setlength{\bibsep}{0.13\baselineskip}
\enlargethispage{2\baselineskip}
\vspace{-1em}
\renewcommand{\refname}{%
\vspace{-1em}
 References}

\bibliographystyle{JHEP}
\bibliography{bibliography}

%\begin{thebibliography}{99}
%\bibitem{...}
%\end{thebibliography}

\end{document}